\documentclass[conference]{IEEEtran}
\usepackage{epsfig}
\usepackage{amsmath,amssymb,amstext}
\usepackage{epstopdf}
\usepackage{graphicx}

\begin{document}
\title{Beamforming for Magnetic Induction \\based Wireless Power Transfer Systems \\with Multiple Receivers}
\author{S. Kisseleff$^*$, I. F. Akyildiz$^{**}$, and W. Gerstacker$^*$\\
$^*$ Institute for Digital Communications, \\Friedrich-Alexander University (FAU) Erlangen-N\"urnberg, Germany, \{kisseleff, gersta\}@lnt.de\\
$^{**}$ Broadband Wireless Networking Lab, Georgia Institute of Technology, USA, ian@ece.gatech.edu
}
\maketitle
\begin{abstract}
Magnetic induction (MI) based communication and power transfer systems have gained an increased attention in the recent years. Typical applications for these systems lie in the area of wireless charging, near-field communication, and wireless sensor networks. For an optimal system performance, the power efficiency needs to be maximized. Typically, this optimization refers to the impedance matching and tracking of the split-frequencies. However, an important role of magnitude and phase of the input signal has been mostly overlooked. Especially for the wireless power transfer systems with multiple transmitter coils, the optimization of the transmit signals can dramatically improve the power efficiency. In this work, we propose an iterative algorithm for the optimization of the transmit signals for a transmitter with three orthogonal coils and multiple single coil receivers. The proposed scheme significantly outperforms the traditional baseline algorithms in terms of power efficiency. 
\end{abstract}
\section{Introduction}
\label{sec:1}
Magnetic induction (MI) based transmissions are well known in the context of near-field communication (NFC) \cite{near_field_Mag}, wireless power transfer (WPT) \cite{Charge_coils}, and wireless sensor networks (WSN) in challenging environments \cite{MI_comms_WUSN}, \cite{underwater_MI}. In this work, our main focus lies in the WPT using resonant coupling of magnetic antennas. Typically, MI based power transfer is only useful within a short range due to a dramatically low power efficiency otherwise, as it has been confirmed in numerous previous works (e.g. \cite{WPT_book}). Furthermore, the alignment of coils has a strong impact on the transfer efficiency, see e.g. \cite{misalignment}, \cite{Interference_polariz}. Several attempts have been made to extend the magnetic induction based point-to-point transmission to a system with multiple receivers \cite{WPT_MultipleReceivers}, multiple transmitters \cite{WPT_MultipleTransmitters}, or even multiple relays \cite{MI_waveguide_first}, \cite{agbinya_masihpour_relaying}. Moreover, MI based networks with multiple transceivers and relays have been analyzed for the underground WSNs \cite{WUSN_KeyOpt}. Finally, the multiple-input multiple-output (MIMO) technique has been introduced for different constellations of MI based communication and WPT systems, see e.g. \cite{nearfield_MIMO}, \cite{mi_net_MINERS}. In particular, \cite{mi_net_MINERS} suggests the use of a transmitter equipped with three orthogonally deployed coils. This enables a steerable directionality of the resulting magnetic field, the so-called magnetic vector modulation. This decomposition is based on the explicit weighting of the field vectors given by the orientations of the transmitter coils \cite{mi_net_MINERS}. Unfortunately, this approach does not take into account the influence of multiple receivers on each other or on the transmitter. Similarly, in \cite{nearfield_MIMO}, only the carrier frequency is optimized, leaving the choice of e.g. signal phase and amplitude (beamforming coefficients\footnote{We follow the convention of WPT using electromagnetic waves \cite{WPT_book} and adopt the term beamforming for the optimization of the transmit signal vector in spatial domain.}) suboptimal. \\
Beamforming is a well known technique for maximizing the power efficiency of a MIMO system. In traditional radio frequency (RF) systems, the power efficiency optimization corresponds to the maximization of the receive power for fixed $L_2$-norm of the beamforming vector, because the consumed power in the transmitting device depends only on the constraint on $L_2$-norm of the beamforming vector, not on its particular coefficients. For MI based WPT, the efficiency depends explicitly on the coupling between the transceivers, such that improving the coupling yields an increase of the power efficiency \cite{Charge_coils}. Furthermore, since the signal reflection is approximately proportional to the squared mutual inductance \cite{chest_MI_2014}, an influence of the reflected signals on the transmit power is inevitable. In particular, these reflected signals can overlap constructively or destructively depending on the phase of the input signals. Hence, the transmit power depends on the choice of the beamforming coefficients and a sole receive power maximization becomes insufficient. Therefore, we propose an iterative algorithm, which takes into account all couplings between the coils and maximizes the WPT efficiency.\\
For this work, we consider one transmitter with three ortho-gonal coils free of self-interference and multiple single antenna receivers randomly deployed in the near-field of the transmitter. For a more flexible system design, the power transfer links can be assigned different priorities, such that more power can be steered into the preferred direction. In this context, high efficiency gains can be observed compared to the mentioned baseline schemes. In addition, a WPT efficiency of up to $96\%$ can be achieved even for more than three receivers.\\
This paper is organized as follows. Section \ref{sec:2} provides insight into the system model for WPT with three transmission coils and multiple receivers. Our system model  allows also a priority based optimization, such that some receivers may get more power upon request. In Section \ref{sec:3}, the efficiency optimization problem is formulated and different solutions are introduced. In Section \ref{sec:4}, numerical results are provided, and Section \ref{sec:5} concludes the paper.
\section{System Model}
\label{sec:2}
\begin{figure}
\centering
\includegraphics[width=2.2in]{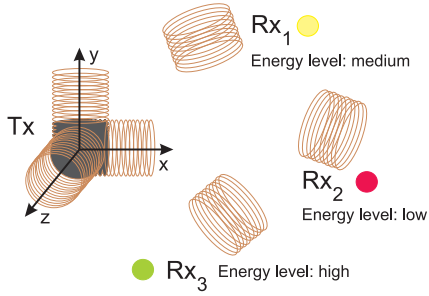}
\caption{Example of a WPT system with 3D-coil based transmitter ($Tx$) and multiple ($K=3$) single coil receivers ($Rx_1$, $Rx_2$, and $Rx_3$). Different energy levels motivate the priority aware efficiency optimization.}
\label{network}
\end{figure}

\begin{figure}
\centering
\includegraphics[width=2.1in]{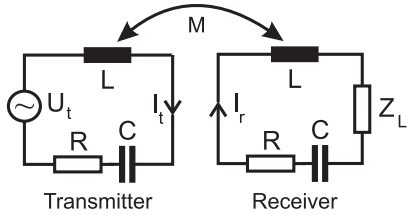}
\caption{A single WPT link. Transmitter and receiver resonance circuits.}
\label{circuits}
\vspace*{-2mm}
\end{figure}

In this work, we utilize one transmitter equipped with three orthogonally deployed coils (3D-coil) and multiple ($K$) receivers with one coil each, see Fig. \ref{network}, where $K=3$. 
Every coil with inductivity $L$ is considered as part of a resonant circuit, which includes also a capacitor with capacitance $C$ and a resistor with resistance $R$ (modeling the copper resistance of the coil). The capacitance $C$ is selected to make the circuit resonant at the resonance frequency $f_0=\frac{1}{2\pi\sqrt{LC}}$. However, the actual operating frequency\footnote{For the WPT, only one frequency is utilized \cite{WPT_book}.} $f$ for the WPT can be subject to optimization. Furthermore, each receiver circuit contains a real-valued load resistor $Z_L$, which can be optimized individually for different receivers in order to minimize the power reflection at the receiver according to the previous work \cite{Charge_coils}, see Fig. \ref{circuits}.
Hence, the inner impedance of a resonant circuit with index $k$ can be given by  
\begin{equation}
\label{eq:2_1}
Z_{k, k}=j2\pi fL+\frac{1}{j2\pi fC}+R,
\end{equation}
if circuit $k$ belongs to the transmitter, or
\begin{equation}
\label{eq:2_2}
Z_{k, k}=j2\pi fL+\frac{1}{j2\pi fC}+R+Z_{L, k},
\end{equation}
if circuit $k$ belongs to a receiver. The induced voltage is related to the coupling between the coils, which is determined by the mutual inductance $M$. The knowledge of the mutual inductance is very important for the optimization of WPT systems \cite{Charge_coils} and can be determined either by channel estimation or by distance estimation under the assumption that all other system parameters (coil dimensions, polarization, etc.) are known. In this work, we assume that the exact value of the mutual inductance between any pair of coils is available to the transmitter. With this, the frequency selectivity of the MI channel is entirely known, such that not only the WPT can be established, but possibly also the information transmission. For this, only the signal pulse shape needs to be adapted\footnote{For information transmission, bandlimited signals like root-raised cosine pulses are typically utilized. For WPT, the transmitted signal is usually a sine wave.}. However, MI based information transmission is beyond the scope of this work.\\
The orientations and alignments of the coupled coils have a strong impact on $M$ and correspondingly on the path loss, cf. \cite{Interference_polariz}. Hence, we model the mutual inductance between coils $k$ and $l$ by
\begin{eqnarray}
\label{eq:2_3_1}
M_{k, l}\hspace{-2mm}&=&\hspace{-2mm}\overline{M}_{k, l}\cdot J_{k, l},\\
\label{eq:2_3_2}
J_{k, l}\hspace{-2mm}&=&\hspace{-2mm}2\sin{\theta_k}\sin{\theta_l}+\cos{\theta_k}\cos{\theta_l}\cos{\phi},
\end{eqnarray}
cf. \cite{Interference_polariz}, where $\theta_k$ and $\theta_l$ are the angles between the radial directions of the coils $k$ and $l$, respectively, and the line connecting the two coil centers. $\phi$ is the angle difference between the coils' axes in the plane, which is orthogonal to the direction of transmission. $\overline{M}_{k, l}$ represents the (absolute) value of the mutual inductance for the case $J_{k, l}=1$. For the following, we define $Z_{k, l}=j2\pi fM_{k, l}, \:\forall k\neq l$.\\
We consider the complex-valued amplitudes $U_k$ and $I_k$ of the voltages $u_k(t)=U_k\cdot\operatorname{e}^{j2\pi ft}$ and currents $i_k(t)=I_k\cdot\operatorname{e}^{j2\pi ft}, \: \forall k$, respectively. For each coil $k$, the current amplitude $I_{k}$ in the resonant circuit depends on the current amplitudes $I_l, \:\forall l\neq k$ in all surrounding circuits via the voltage equation
\begin{equation}
\label{eq:2_4}
I_{k}\cdot Z_{k, k}+\sum_{l\neq k}\left(I_l\cdot Z_{k, l}\right)=U_k,
\end{equation} 
where $U_k$ is the complex-valued amplitude of the input voltage. In the following, the superscripts $(\cdot)^T$ and $(\cdot)^H$ denote transpose and Hermitian transpose, respectively. We assign the first three coil indices to the transmitter and the remaining indices to the receivers. 
In order to calculate the currents in all circuits of the coupled network, a set of voltage equations
\begin{eqnarray}
\label{eq:2_6}
\begin{bmatrix}
\textbf{Z}_{Tx} & \textbf{Z}_{Ch}\\[1mm]
\textbf{Z}_{Ch}^T & \textbf{Z}_{Rx}
\end{bmatrix}\cdot
\begin{bmatrix}
\textbf{I}_{Tx}^c\\[1mm]
\textbf{I}_{Rx}^c
\end{bmatrix}=
\begin{bmatrix}
\textbf{U}_{Tx}\\[1mm]
\textbf{0}
\end{bmatrix}
\end{eqnarray}
needs to be solved. Here, $\textbf{U}_{Tx}$ is the complex-valued input voltage vector at the transmitter. Furthermore, $\textbf{0}$ stands for the all-zero vector and represents the input voltages $u_k(t)\equiv0$ for all coils $k$ belonging to a receiver, since only the transmitter is supposed to generate power. $\textbf{I}^c_{Tx}$ and $\textbf{I}^c_{Rx}$ denote the current vectors of the transmitter and the receiver circuits, respectively. The matrices $\textbf{Z}_{Tx}$, $\textbf{Z}_{Ch}$ and $\textbf{Z}_{Rx}$ contain complex impedances and are defined in the following. In this work, we use a 3D-coil based transmitter, which means that all three transmitter coils' axes are orthogonal to each other, such that
\begin{eqnarray}
\label{eq:2_7}
\textbf{Z}_{Tx}=
\begin{bmatrix}
Z_{1, 1} & 0 & 0\\
0 & Z_{2, 2} & 0\\
0 & 0 & Z_{3, 3}
\end{bmatrix}
\end{eqnarray}
holds. The receiver coils are not necessarily orthogonal and we obtain
\begin{eqnarray}
\label{eq:2_7_1}
\textbf{Z}_{Rx}=
\begin{bmatrix}
Z_{4, 4} & \cdots & Z_{K+3, 4}\\
\vdots & \ddots & \vdots\\
Z_{4, K+3} & \cdots & Z_{K+3, K+3}
\end{bmatrix}.
\end{eqnarray}
The purely imaginary matrices $\textbf{Z}_{Ch}$ and $\textbf{Z}_{Ch}^T$ in \eqref{eq:2_6} stand for the influence of the receiver coils onto the transmitter coils and vice versa, respectively. Hence, $\textbf{Z}_{Ch}$ is defined by
\begin{eqnarray}
\label{eq:2_7_1}
\textbf{Z}_{Ch}=
\begin{bmatrix}
Z_{4, 1} & Z_{5, 1} & \cdots & Z_{K+3, 1}\\
Z_{4, 2} & Z_{5, 2} & \cdots & Z_{K+3, 2}\\
Z_{4, 3} & Z_{5, 3} & \cdots & Z_{K+3, 3}
\end{bmatrix}.
\end{eqnarray}
By inverting the impedance matrix in \eqref{eq:2_6} using \cite{BlockMatrix_book}, we obtain similar to \cite{WPT_MultipleReceivers}
\begin{eqnarray}
\label{eq:2_8_1}
\textbf{I}_{Tx}^c\hspace*{-2mm}&=&\hspace*{-2mm}\left(\textbf{Z}_{Tx}-\textbf{Z}_{Ch}\textbf{Z}_{Rx}^{-1}\textbf{Z}_{Ch}^T\right)^{-1}\textbf{U}_{Tx}=\textbf{A}\textbf{U}_{Tx},\\
\label{eq:2_8_2}
\textbf{I}_{Rx}^c\hspace*{-2mm}&=&\hspace*{-2mm}-\textbf{Z}_{Rx}^{-1}\textbf{Z}_{Ch}^T\textbf{A}\textbf{U}_{Tx}=\textbf{C}\textbf{U}_{Tx},
\end{eqnarray}
with implicit definitions of $\textbf{A}$ and $\textbf{C}$. As known from the fundamentals of electric power generation and transmission (e.g. \cite{electric_book}, \cite{electric_book2}), in order to produce enough active power in electric circuits, the transmitter/generator needs to release also the reactive power, which corresponds to the imaginary part of the generated complex power. The reactive power is not absorbed by the load, but fluctuates between the power source and the load impedance. Furthermore, without the reactive power, the induction coils cannot be operated. Hence, both real and imaginary parts need to be taken into account, such that the magnitude of the generated complex power (the so-called apparent power) is a better reference for the maximum transmit power than the pure active power \cite{electric_book2}. In the $k$th transmitter circuit, the apparent power is given by
\begin{equation}
\label{eq:2_9}
P_{t, k}=\left|U_{k}I_{k}\right|=\left|U_{k}\right|\left|I_{k}\right|,
\end{equation}
where $\left|\cdot\right|$ denotes the element-wise absolute value operator. Therefore, we define the total power provided by the transmitter as
\begin{equation}
\label{eq:2_10}
P_{t, \mathrm{total}}=\sum_{k=1}^3 P_{t, k}=\left|\textbf{U}_{Tx}\right|^T\left|\textbf{A}\textbf{U}_{Tx}\right|.
\end{equation}
For the received active power at the load resistor $Z_{L, l}$ of the receiver circuit $l$ we obtain
\begin{equation}
\label{eq:2_11}
P_{r, l}=\left|I_l\right|^2Z_{L, l}.
\end{equation}
In practical WPT systems (in particular related to the sensor networks), the energy consumption in receiver devices may be not the same, such that the depletion rate of the batteries varies. Therefore, some receiver devices may require more power than the others, see Fig. \ref{network}. Hence, we introduce the priority coefficients $W_l, \:\forall l$. With this, the weighted received power is given by
\begin{eqnarray}
P_{r, \mathrm{total}}\hspace{-2mm}&=&\hspace{-2mm}\sum_{l=4}^{K+3}W_lP_{r, l}\notag\\
\label{eq:2_13}
&&\hspace{-20mm}=\textbf{U}_{Tx}^H\textbf{C}^H
\begin{bmatrix}
W_4Z_{L, 4} & \cdots & 0\\
\vdots & \ddots & \vdots\\
0 & \cdots & W_{K+3}Z_{L, K+3}
\end{bmatrix}
\textbf{C}\:\textbf{U}_{Tx},
\end{eqnarray}
where \eqref{eq:2_8_2} and \eqref{eq:2_11} have been used.
\section{Beamforming}
\label{sec:3}
In this section, the design of the optimal input vector $\textbf{U}_{Tx}$ (MI beamforming vector) for the weighted power efficiency maximization is discussed. First, the optimization problem is formulated. Then, some of the most promising approaches including the proposed iterative algorithm are presented.
\subsection{Problem formulation}
\label{sec:3_1}
In this work, we define the WPT efficiency with respect to the sum apparent power given by \eqref{eq:2_10}. Hence, the optimization problem can be formulated as
\begin{equation}
\label{eq:3_1}
\max_{\hspace*{-2mm}\textbf{U}_{Tx}, f, Z_{L, l}\forall l}\frac{P_{r, \mathrm{total}}}{P_{t, \mathrm{total}}},
\end{equation}
where the parameters $f$ and $Z_{L, l},\:\forall l$ can be optimized according to the literature \cite{WPT_MultipleReceivers}, \cite{nearfield_MIMO}. Hence, we focus on the optimization of $\textbf{U}_{Tx}$.\\
It can be shown that the problem \eqref{eq:3_1} is non-convex due to the non-convexity of $P_{t, \mathrm{total}}$, such that the well-known convex optimization tools \cite{bconvex} cannot be used. Therefore, we provide some suboptimal schemes. Our proposed algorithm discussed in Section \ref{sec:3_2_3} can be shown to reach a local optimum in case of convergence.
\subsection{Proposed algorithms}
\label{sec:3_2}
In the following, three different approaches are described.
\subsubsection{Closest neighbor based beamforming}
\label{sec:3_2_1}
As known from the near-field communication and coupled-mode theory, usually only the closest neighbors tend to establish a strong coupling, such that the path losses between the transmitter and any other receiver are dramatically larger, especially in case of weak couplings between coils. Based on this principle, the idea of optimizing the beamforming vector for the closest neighbor of the transmitter is motivated. Hence, using \eqref{eq:2_3_1}-\eqref{eq:2_4}, and assuming $\overline{M}_{l, 1}=\overline{M}_{l, 2}=\overline{M}_{l, 3}$, the magnetic induction at the receiver $l$ is related to
\begin{equation}
\label{eq:3_2}
U_l\propto\left[I_1, I_2, I_3\right]\left[J_{l, 1}, J_{l, 2}, J_{l, 3}\right]^T\cdot j2\pi f\overline{M}_{l, 1},
\end{equation}
where the current contributions from other receiver coils are neglected due to a very high path loss of such signals\footnote{Since here the target receiver $l$ is the closest receiver to the transmitter, the signals arriving from the neighboring coils are basically heavily attenuated reflections from the receivers that are further away from the transmitter.}.
Furthermore, due to identical sets of circuit elements in all transmitter resonance circuits, $I_k\propto U_k, \: k=\{1, 2, 3\}$ holds, which results directly from \eqref{eq:2_7} and \eqref{eq:2_8_1} for weakly coupled coils. Hence, we substitute this result in \eqref{eq:3_2} and obtain
\begin{equation}
\label{eq:3_3}
U_l\propto\left[U_1, U_2, U_3\right]\left[J_{l, 1}, J_{l, 2}, J_{l, 3}\right]^T.
\end{equation}
Due to \eqref{eq:3_3}, the beamforming vector
\begin{equation}
\label{eq:3_4}
\textbf{U}_{Tx}=\left[J_{l, 1}, J_{l, 2}, J_{l, 3}\right]^T\cdot \mathrm{(1\: V)}
\end{equation}
corresponds to the maximum ratio combining (MRC) solution. Obviously, this approach provides a close-to-optimum solution in case of a weak coupling between any adjacent coils and with a dominant receiver. Moreover, any variations of the matrices $\textbf{A}$ and $\textbf{C}$ due to a stronger coupling between neighboring coils result in a deviation of the optimum solution from this beamforming solution, thus decreasing the power efficiency. In addition, the transmit power is not taken into account at all, such that the power efficiency of this solution is suboptimal.
\subsubsection{Receive power maximization}
\label{sec:3_2_2}
The second approach is based on the eigenvalue decomposition of the receiver matrix
\begin{equation}
\label{eq:3_5}
\textbf{D}=\textbf{C}^H
\begin{bmatrix}
W_4Z_{L, 4} & \cdots & 0\\
\vdots & \ddots & \vdots\\
0 & \cdots & W_{K+3}Z_{L, K+3}
\end{bmatrix}
\textbf{C}
\end{equation}
and corresponds to a typical beamforming solution in the traditional RF systems. The eigenvector pertaining to the maximum eigenvalue maximizes the total receive power \eqref{eq:2_13}. This approach takes into account the impact of the coupling of coils onto the receive power. Hence, it provides a more accurate solution for the receive power maximization. \\
For weak couplings between coils (low mutual inductance) and identical inner impedances $Z_{1, 1}=Z_{2, 2}=Z_{3, 3}$ in the transmitter coils, matrix $\textbf{A}$ is approximately\footnote{For $\displaystyle\left(2\pi f\max_{k, l}\{\overline{M}_{k, l}\}\right)\ll R$, $\textbf{Z}_{Ch}^T\textbf{Z}_{Rx}^{-1}\textbf{Z}_{Ch}$ in \eqref{eq:2_8_1} is negligible.} given by
\begin{equation}
\label{eq:3_6}
\textbf{A}\approx\textbf{Z}_{Tx}^{-1}=\left|Z_{1, 1}\right|^{-1}\cdot\textbf{I},
\end{equation}
where $\textbf{I}$ is the identity matrix. The correctness of \eqref{eq:3_6} is confirmed by previous works, e.g. \cite{chest_MI_2014}. This yields
\begin{equation}
\label{eq:3_7}
P_{t, \mathrm{total}}\approx\textbf{U}_{Tx}^H\textbf{U}_{Tx}\cdot \left|Z_{1, 1}\right|^{-1},
\end{equation}
where $\left|\textbf{U}_{Tx}\right|^T\left|\textbf{U}_{Tx}\right|=\textbf{U}_{Tx}^H\textbf{U}_{Tx}$ has been used in \eqref{eq:2_10}.
Therefore, the beamforming optimization problem reduces to an eigenvalue problem given by
\begin{equation}
\label{eq:3_7_2}
\max_{\textbf{U}_{Tx}}\frac{\textbf{U}_{Tx}^H\textbf{D}\:\textbf{U}_{Tx}}{\textbf{U}_{Tx}^H\textbf{U}_{Tx}}.
\end{equation}
Hence, this approach is optimal for the weak couplings between coils. With increasing mutual inductance, the approximations \eqref{eq:3_6} and \eqref{eq:3_7} are not valid anymore. Correspondingly, the efficiency \eqref{eq:3_7_2} becomes suboptimal, because the transmit power according to \eqref{eq:2_10} is not explicitly considered. Thus, a more powerful algorithm is proposed in the following.
\subsubsection{Proposed iterative algorithm}
\label{sec:3_2_3}
The proposed idea is to approximate the total transmit power by a squared $L_2$-norm in each iteration of the algorithm. Using this approximation, the optimal beamforming vector is calculated, which helps updating the solution in the next iteration. For the approximation, we assume that in case of convergence of this algorithm,
\begin{equation}
\label{eq:3_8}
\left|\textbf{U}_{Tx, n}\right|\approx \left|\textbf{U}_{Tx, n-1}\right|
\end{equation}
holds, where $\textbf{U}_{Tx, n}$ denotes the state of the vector $\textbf{U}_{Tx}$ at the end of the $n$th iteration. At first, we approximate the transmit power \eqref{eq:2_10} by
\begin{equation}
\label{eq:3_8_2}
P_{t, \mathrm{total}}=\left|\textbf{U}_{Tx, n}\right|^T\left|\textbf{A}\textbf{U}_{Tx, n}\right|\approx\left|\textbf{U}_{Tx, n-1}\right|^T\left|\textbf{A}\textbf{U}_{Tx, n}\right|,
\end{equation}
such that the order of the transmit power with respect to $\textbf{U}_{Tx, n}$ reduces\footnote{The order with respect to the complex-valued variable $\textbf{U}_{Tx, n}$ is larger for $\left|\textbf{U}_{Tx, n}\right|^T\left|\textbf{A}\textbf{U}_{Tx, n}\right|$ than for $\left|\textbf{A}\textbf{U}_{Tx, n}\right|$.}. Then, we express $\left|\textbf{U}_{Tx, n-1}\right|^T$ as
\begin{equation}
\label{eq:3_8_3}
\left|\textbf{U}_{Tx, n-1}\right|^T=\left[1, 1, 1\right]\textbf{V}_n,
\end{equation}
using matrix $\textbf{V}_n$ defined by
\begin{eqnarray}
\label{eq:3_9}
\textbf{V}_{n}\hspace{-2mm}&=&\hspace{-2mm}
\begin{bmatrix}
\left|U_{1, n-1}\right| & 0 & 0\\
0 & \left|U_{2, n-1}\right| & 0\\
0 & 0 & \left|U_{3, n-1}\right|
\end{bmatrix}.
\end{eqnarray}
By inserting \eqref{eq:3_8_3} into \eqref{eq:3_8_2} and using \eqref{eq:3_9}, we obtain
\begin{equation}
\label{eq:3_9_2}
P_{t, \mathrm{total}}\approx\left[1, 1, 1\right]\textbf{V}_n\left|\textbf{A}\textbf{U}_{Tx, n}\right|=\left[1, 1, 1\right]\left|\textbf{V}_n\textbf{A}\textbf{U}_{Tx, n}\right|.
\end{equation}
Moreover, \eqref{eq:3_9_2} can be transformed into a squared $L_2$-norm. For this, we define $\textbf{S}_n=\textbf{V}_n\textbf{A}$ and approximate $\left|\textbf{V}_n\textbf{A}\textbf{U}_{Tx, n}\right|=\left|\textbf{S}_n\textbf{U}_{Tx, n}\right|$ from \eqref{eq:3_9_2} by
\begin{equation}
\label{eq:3_11}
\left|\textbf{S}_n\textbf{U}_{Tx, n}\right|\approx\left|\textbf{S}_n\textbf{U}_{Tx, n}\right|\otimes\left(\left|\textbf{S}_n\textbf{U}_{Tx, n}\right|\oslash\left|\textbf{S}_n\textbf{U}_{Tx, n-1}\right|\right),
\end{equation}
where $\otimes$ and $\oslash$ represent element-wise vector multiplication and division, respectively. Hence, by reformulating \eqref{eq:3_11}, $\left|\textbf{S}_n\textbf{U}_{Tx, n}\right|$ can be expressed as
\begin{equation}
\label{eq:3_11_2}
\left|\textbf{S}_n\textbf{U}_{Tx, n}\right|\approx\left(\left|\textbf{S}_n\textbf{U}_{Tx, n}\oslash\sqrt{\left|\textbf{S}_n\textbf{U}_{Tx, n-1}\right|}\right|\right)^ {\textcircled{\tiny 2}},
\end{equation}
where $\left(\cdot\right)^{\textcircled{\tiny 2}}$ denotes element-wise square operator. By multiplying \eqref{eq:3_11_2} with a vector $\left[1, 1, 1\right]$, the transmit power can be expressed as a squared $L_2$-norm using \eqref{eq:3_9_2}:
\begin{equation}
\label{eq:3_11_3}
\left|\textbf{U}_{Tx, n}\right|^T\left|\textbf{A}\textbf{U}_{Tx, n}\right|\approx \big{\|}\textbf{S}_n\textbf{U}_{Tx, n}\oslash\sqrt{\left|\textbf{S}_n\textbf{U}_{Tx, n-1}\right|}\big{\|}_2^2.
\end{equation}
For the clarity of exposition, we denote $\left|\textbf{S}_n\textbf{U}_{Tx, n-1}\right|$ by vector $\textbf{G}_n=\left[G_1, G_2, G_3\right]^T$. Using $\textbf{G}_n$, the element-wise division in \eqref{eq:3_11_3} can be formulated as a multiplication with a matrix $\textbf{Q}_n$, where
\begin{eqnarray}
\label{eq:3_12}
\textbf{Q}_{n}\hspace*{-2mm}&=&\hspace*{-2mm}
\begin{bmatrix}
\sqrt{G_1}^{-1} & 0 & 0\\
0 & \sqrt{G_2}^{-1} & 0\\
0 & 0 & \sqrt{G_3}^{-1}
\end{bmatrix}.
\end{eqnarray}
Then, we obtain
\begin{equation}
\label{eq:3_12_2}
\textbf{S}_n\textbf{U}_{Tx, n}\oslash\sqrt{\left|\textbf{S}_n\textbf{U}_{Tx, n-1}\right|}=\textbf{Q}_{n}\textbf{S}_n\textbf{U}_{Tx, n}.
\end{equation}
Finally, by inserting \eqref{eq:3_12_2} into \eqref{eq:3_11_3}, we obtain
\begin{equation}
\label{eq:3_13}
\left|\textbf{U}_{Tx, n}\right|^T\left|\textbf{A}\textbf{U}_{Tx, n}\right|\approx\|\textbf{Q}_{n}\textbf{S}_n\textbf{U}_{Tx, n}\|_2^2.
\end{equation}
Based on \eqref{eq:3_13}, the beamforming problem \eqref{eq:3_1} can be reduced to a generalized eigenvalue problem in each iteration
\begin{equation}
\label{eq:3_13_2}
\textbf{U}_{Tx, n}=\arg\max_{\hspace*{-4mm}\textbf{U}_{Tx}}\frac{\textbf{U}_{Tx}^H\textbf{D}\:\textbf{U}_{Tx}}{\textbf{U}_{Tx}^H\left(\textbf{Q}_{n}\textbf{S}_n\right)^H\left(\textbf{Q}_{n}\textbf{S}_n\right)\textbf{U}_{Tx}},
\end{equation}
where $\textbf{D}$ is given by \eqref{eq:3_5}. The solution to this problem is typically computed using a substitution 
\begin{eqnarray}
\label{eq:3_14}
\textbf{X}_n\hspace*{-2mm}&=&\hspace*{-2mm}\left(\textbf{Q}_{n}\textbf{S}_n\right)\textbf{U}_{Tx},\\
\label{eq:3_15}
\textbf{U}_{Tx}\hspace*{-2mm}&=&\hspace*{-2mm}\left(\textbf{Q}_{n}\textbf{S}_n\right)^{-1}\textbf{X}_n.
\end{eqnarray} 
An eigenvalue decomposition is applied to the matrix $\left(\left(\textbf{Q}_{n}\textbf{S}_n\right)^{-1}\right)^H\textbf{D}\left(\left(\textbf{Q}_{n}\textbf{S}_n\right)^{-1}\right)$ and the eigenvector with the maximum eigenvalue is picked as the optimal solution for $\textbf{X}_n$. Then, using \eqref{eq:3_15}, the optimal beamforming vector $\textbf{U}_{Tx, n}$ is calculated. This vector replaces $\textbf{U}_{Tx, n-1}$ in the next iteration. For the starting point $\textbf{U}_{Tx, 0}$, we choose the eigenvector according to Section \ref{sec:3_2_2}. Hence, we start with the receive power maximization without taking into account the influence of matrix $\left(\textbf{Q}_n\textbf{S}_n\right)$ and then improve the energy efficiency using the proposed approach. For a stopping condition, a maximum number of iterations or a minimum efficiency gain of the current iteration over the previous iteration can be used.\\
The major benefit of the proposed algorithm is due to the adaptation of the beamforming to the changes of the transmit power. In case of convergence, the approximations \eqref{eq:3_8}, \eqref{eq:3_11}, and correspondingly \eqref{eq:3_13} are valid. Then, the maximization problem in \eqref{eq:3_1} becomes concave, and the equivalent minimization problem becomes convex \cite{bconvex} and can be solved using the generalized eigenvalue decomposition. Thus, the algorithm leads to a locally optimal solution in case of convergence. Unfortunately, the existence or absence of any other locally optimum solutions cannot be shown mathematically, such that the obtained solution is not necessarily globally optimal. However, as shown in Section \ref{sec:4}, the proposed solution performs well and reaches high power efficiencies.
\section{Numerical Results}
\label{sec:4}
\begin{figure}
\centering
\includegraphics[width=0.48\textwidth]{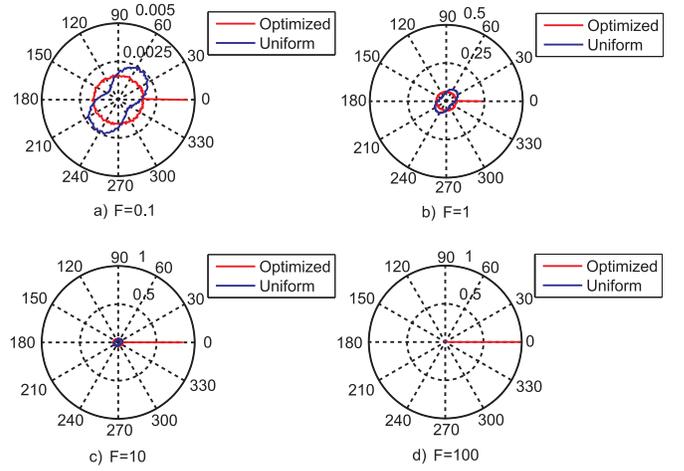}
\caption{Beamforming pattern for $F=\{0.1, 1, 10, 100\}$.}
\label{Beamforming}
\vspace*{-2mm}
\end{figure}

\begin{figure}
\centering
\includegraphics[width=0.48\textwidth]{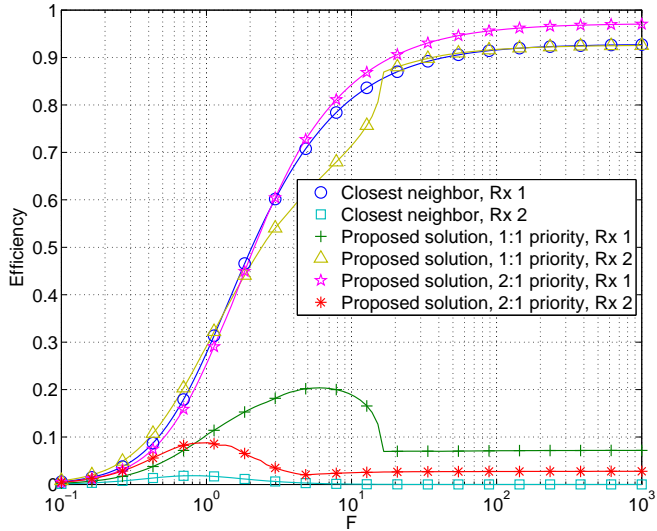}
\caption{Example on power efficiencies versus $F$ of two randomly deployed receivers using different priority metrics.}
\label{Priority_2Rx}
\vspace*{-2mm}
\end{figure}

In this section, we present numerical results for the MI beamforming optimization. In order to follow the convention of the WPT community \cite{WPT_book}, we consider a factor $F_{k, l}=\frac{2\pi f\overline{M}_{k, l}}{R}$, which corresponds to the product of the quality factor $\frac{2\pi fL}{R}$ and the coupling coefficient $\frac{\overline{M}_{k, l}}{L}$ between coils $k$ and $l$, respectively, in our performance investigations. We assume that all receivers are placed at the same distance $d$ from the transmitter, such that the coupling coefficient is identical for all transmitter-receiver links, which means $\overline{M}_{k, l}=\overline{M}, \:k\in\{1, 2, 3\}, \:l\in\{4, \ldots, K+3\}$ and $F_{k, l}=F, \:k\in\{1, 2, 3\}, \:l\in\{4, \ldots, K+3\}$. Since the distances between the receivers may vary, the mutual inductance between them differs from $\overline{M}$. Assuming that the distance between two adjacent receivers with indices $l_1$ and $l_2$ is $d_{l_1, l_2}$, $\overline{M}_{l_1, l_2}$ can be expressed as
\vspace*{-1mm}
\begin{equation}
\overline{M}_{l_1, l_2}=\overline{M}\left(\frac{d}{d_{l_1, l_2}}\right)^3,
\end{equation} 
because the mutual inductance scales with the third power of the transmission distance \cite{near_field_Mag}. For simplicity, the carrier frequency is selected to be equal to the resonance frequency, $f=f_0$.\footnote{In principle, the optimal carrier frequency according to \cite{nearfield_MIMO} could be selected.} In order to provide insight into practically relevant values for the factor $F$, we calculate it for the following scenario. Assume, a single turn rectangular air core coil of 4 cm $\times$ 6 cm cross-section area made from copper wire of 3 mm thickness is operated at $f_0=125$ MHz. Then, for a transmission distance $d=0.4$ m, we obtain $F\approx 15$ \cite{Friedrich_table}.\\
We start with the visualization of the beamforming for a single receiver.  For this, we assume that the receiver with randomly rotated coil is at first placed at $0^\circ$  of the angular space of the transmitter and the optimal transmit signal $\textbf{U}_{Tx, \mathrm{optimized}}$ is found using the proposed algorithm. This signal is then used in order to evaluate the beamforming effect of this particular constellation. For this, the receiver is moved around the transmitter and its axis orientation is randomly rotated. For each point in the angular space we determine the power efficiency using $\textbf{U}_{Tx, \mathrm{optimized}}$. This calculation is repeated for 1000 different constellations. The mean value of the resulting efficiency pattern is shown in Fig. \ref{Beamforming}.
For comparison, we also show the results for a uniform beamforming vector $\textbf{U}_{Tx, \mathrm{uniform}}=[1, 1, 1]^T\cdot (1\: V)$. This baseline scheme is motivated by \cite{mi_net_MINERS}, where it is suggested to maximize the field strength in the preferred direction by simply increasing the transmit power in the circuit with corresponding coil orientation. On the contrary, using $\textbf{U}_{Tx, \mathrm{uniform}}$, a quasi-omnidirectional field propagation is supposed to be achieved. As we can see from Figs. \ref{Beamforming}a) and b), this field propagation is on average not omnidirectional and provides a better efficiency for the receivers deployed at $45^\circ$ and $225^\circ$, respectively. The power efficiency of the proposed solution becomes more and more directional with increasing $F$ and converges to a single peak. Hence, the power is steered only into the direction of the dedicated user, which is very beneficial, since almost no power is lost in the non-preferred directions and very limited interference is imposed on the other MI based communication systems. \\
In order to visualize how the priority aware efficiency maximization affects the receive powers, we show an example on the power efficiencies of two receivers for different priority metrics, see Fig. \ref{Priority_2Rx}. These receivers are located at randomly selected positions on the circle in distance $d$ from the transmitter and their coils are randomly rotated. For comparison, also the results of the closest neighbor based optimization are presented, where always the first receiver (Rx 1) is selected as the closest neighbor due to the equal distance of both receivers to the transmitter.
Obviously, for equal priority of the receivers, the second receiver (Rx 2) obtains more power than the first receiver using the proposed solution. The sum of their efficiencies is obviously also larger than that of the closest neighbor based approach. By prioritizing the first receiver over the second (2:1 priority), the receive power at Rx 1 becomes significantly larger. Interestingly, this efficiency even is superior to that of the closest neighbor based optimization, which is supposed to maximize the receive power for Rx 1. This is due to the couplings between coils and the transmit power that have been not taken into account in this approach. For this particular example, the priority increase by factor 2 is already sufficient in order to steer the power into the direction of Rx 1. In general, larger priority factors (10:1 or even 20:1) may be needed in order to obtain the desired performance of the beamforming.\\
Finally, we show the mean achievable power efficiency for WPT to multiple receivers using the algorithms described in this work. For this, we consider five receivers and set all priorities to 1, such that no scaling of the receive powers is performed. The mean power efficiency is then calculated by averaging over the efficiencies from 1000 different constellations, where each constellation corresponds to the realization of a random placement of the given number of receivers in distance $d$ around the transmitter and a random orientation of their coils. Also, for each constellation, the orientation of the transmitter is randomly selected.
\begin{figure}
\centering
\includegraphics[width=0.48\textwidth]{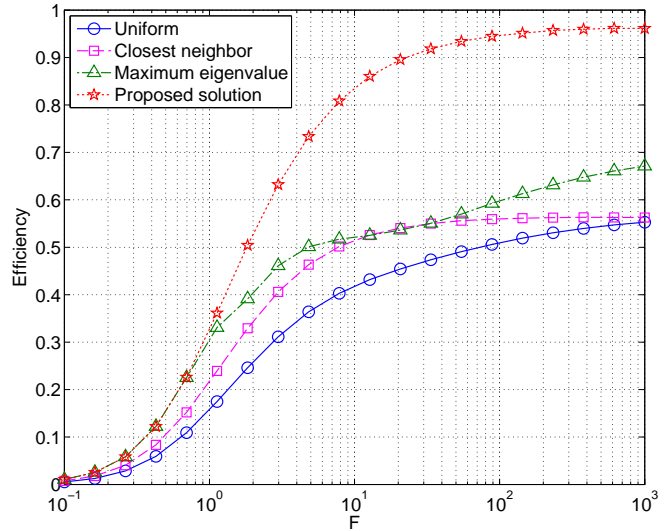}
\caption{Average power efficiency versus $F$ of WPT to five randomly distributed receivers using different algorithms.}
\label{Efficiency_5Rx}
\vspace*{-2mm}
\end{figure}
The results for the average power efficiency versus $F$ depicted in Fig. \ref{Efficiency_5Rx} show significant efficiency improvements of up to $37\%$ for the proposed algorithm compared to the other solutions. As discussed in Section \ref{sec:3_2_1}, the closest neighbor based approach does not take into account the couplings between receivers and the transmit power variations, such that it mostly performs worse or equal to the maximum eigenvalue based approach. As mentioned in Section \ref{sec:3_2_2}, the maximum eigenvalue based beamforming is optimal for weak coupling, therefore its efficiency curve overlaps with that of the proposed solution for $F\leq 1$. For stronger couplings ($F>1$), the proposed solution shows a much steeper increase of the efficiency, yielding large efficiency gains. The mean power efficiency reaches values of 0.96 for large $F$, which corresponds to a power loss of only $4\%$. This makes the proposed algorithm very promising.
\section{Conclusion}
\label{sec:5}
In this work, a novel beamforming solution for the efficiency maximization problem in MI based WPT systems is presented. Due to the non-convexity of the transmit power metric, the optimum solution cannot be determined analytically. Therefore, we presented three suboptimal approaches and discussed their differences. The third approach corresponds to our proposed solution. It utilizes several assumptions, that reduce the order of the optimization problem. In the convergence point of the algorithm, a locally optimal solution for this problem is obtained. Furthermore, a priority aware optimization is possible, that allows for allocation of the power to different receivers according to their needs. For the WPT to multiple receivers, significant gains have been observed, which makes the use of our algorithm very promising.
\bibliographystyle{IEEEtran}
\bibliography{Literatur}
\end{document}